\documentclass[twocolumn,showpacs,showkeys,preprintnumbers,amsmath,amssymb]{revtex4}

\usepackage{graphicx}
\usepackage{dcolumn}
\usepackage{bm}

\usepackage{booktabs}
\usepackage{amsmath}
\usepackage{epstopdf}

\def\beq{\begin{equation}}
\def\eeq{\end{equation}}

\begin{document}

\preprint{DRAFT}

\title{Study of microbunching instability in the linac of a soft X-ray FEL Facility}

\author{Dazhang Huang}
\email{huangdazhang@sinap.ac.cn}
\author{Qiang Gu}%
 \email{guqiang@sinap.ac.cn}
\author{Zhen Wang}
\author{Meng Zhang}
\affiliation{%
Shanghai Institute of Applied Physics, Chinese Academy of Sciences, Shanghai, 201800, China P.R.\\
}%

\author{King Yuen Ng}
\affiliation{
Fermi National Accelerator Laboratory, Batavia, Illinois 60510, USA\\
}

\date{\today}

\begin{abstract}

The development of the microbunching instability is studied for the linac of the proposed Shanghai Soft X-ray Free-Electron-Laser facility (SXFEL) by analytic formulae as well as numerical simulations with the aid of two different codes. The process is investigated in detail and the growth rates (gains) of the instability under various conditions are compared. The results indicate that the limitations from numerical computations in the present simulation model must be taken into account. Moreover, the appearance of higher-order mode excitations in the simulations suggests that further improvement of the current theory is necessary. A mechanism of introducing shot noise into beam profile as the beam passes through a chicane is proposed. The key issues that drive the instability are analyzed.     
\end{abstract}

\maketitle

\section{\label{sec:level1}Introduction}

The microbunching instability in the linac of a free-electron Laser (FEL) facility has always been a problem that degrades the quality of electron beams. The instability is driven by various effects, such as the longitudinal space charge (LSC)~\cite{Saldin}, coherent synchrotron radiation (CSR)~\cite{Heifets,ZHuang1}, and linac wakefields. As the beam passes through a bunch compressor, (e.g., a magnetic chicane) the energy modulation introduced by those effects is transformed into density modulation and thus the instability develops. In a FEL facility, there is usually more than one bunch compressor and the overall growth of the instability is the product of all the gains in each compressor. As a result, the final gain of the instability can become significant. On the other hand, the FEL process has a high demand for electron beam quality in terms of peak current, emittance, energy spread, etc. Therefore without effective control, the microbunching instability can damage the beam quality so seriously that the whole FEL facility fails. Thus learning how to control and/or reduce the instability is a key to the success of a FEL project. 

The Shanghai Soft X-ray FEL Facility (SXFEL) project has been proposed and the feasibility study has been finished. The facility is expected to be built in a few years. Once it is built, it will be the first X-ray FEL facility in China. It is a cascading high-gain harmonic generation (HGHG) FEL facility working at the wavelength of 9~nm in the soft-X-ray region. The basic beam parameters at the exit of linac are: beam energy 840 MeV, peak current 600~A, and normalized transverse emittance less than 2.0~mm.mrad. Both analytical and numerical studies show that the microbunching-instability-induced growth of the global and slice energy spread in the linac is not ignorable without proper control. Moreover, the instability reduces the smoothness of the longitudinal beam current profile as well. One way to control the instability is to increase the uncorrelated energy spread of the beam by a laser heater~\cite{ZHuang}, which will be implemented in the SXFEL.

In this article, we study the microbunching instability based on the design parameters of the SXFEL in order to gain better understanding of the instability process so as to uncover the limitations in analytic formulae and numerical simulations. We realize that the higher harmonics of beam current can be excited in a bunch compressor which may lead to modification to the linear theory of the instability gain. Numerical parameters, such as the half-width and the order of the Savitzky-Golay filter~\cite{Savitzky} employed in the computation of the longitudinal space-charge (LSC) force and coherent-synchrotron-radiation (CSR) effect, play an important role in the energy spread of the local slice. Moreover, the shot noise introduced inevitably in a bunch compressor due to transverse-longitudinal coupling will be important in a laser heater. In Sec.~II, we introduce the basic structure of the SXFEL linac, the fundamental theory of microbunching instability, and the gain curves of the instability in the SXFEL linac obtained by analytic formulae and numerical simulations based on the present design parameters. Comparisons of the gain curves under different conditions are provided and some sensitive parameters are analyzed. In Sec.~III, some critical issues found in the instability study which may have significant effects on the final results are discussed. Summaries and concluding remarks are given in Sec.~IV.       

\section{microbunching instability analysis in SXFEL}

The basic mechanism of the microbunching instability driven by various effects has been studied and corresponding theories have been derived in Refs.~\onlinecite{Saldin,ZHuang,Heifets}. The development of the instability is similar to the amplification process in a klystron amplifier. The initial density modulation and/or white noise are transformed into energy modulation by various kinds of impedance (e.g., LSC, CSR, linac wakefields) during beam transportation. The energy modulation accumulates and is fed-back into the beam after passing through a dispersive section such as a bunch compressor. This accumulation results in much stronger amplified density modulation. Additionally, the CSR effect in the dispersive section forms a positive feedback to enhance the instability. More dispersive sections in a linac result in stronger instability growth. 

\begin{figure}[htb]
   \centering
   \includegraphics*[width=80mm]{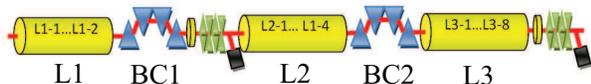}
\vskip-0.05in
   \caption{(Color) Layout of the SXFEL linac.}
   \label{linac}
\end{figure}

\begin{figure}[htb]
   \centering
   \includegraphics*[width=60mm]{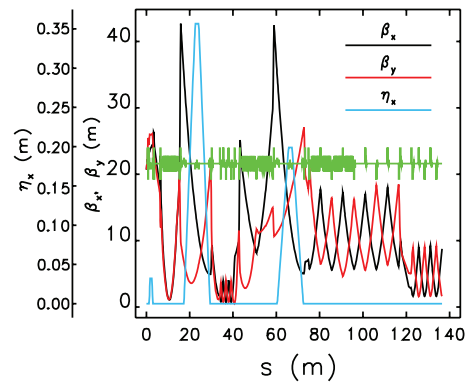}
\vskip-0.05in
   \caption{(Color) Magnet layout and Twiss parameters of the SXFEL linac.}
   \label{magnet}
\end{figure}

In the linac of the SXFEL, both S-band and C-band rf structures are used to accelerate the electron beam. One X-band structure is implemented to suppress the second-order nonlinear components in the longitudinal phase space to avoid undesired growth of transverse emittance and slice energy spread. Two magnetic chicane-type bunch compressors (BC1 and BC2) are used to compress the beam to arrive at the required peak current. The layout of the SXFEL linac is shown in Fig.~\ref{linac}, and the magnet layout and some of the Twiss parameters of the SXFEL linac are shown in Fig.~\ref{magnet}.  Note that S-band accelerating structures are used in sections L1 and L2, whereas a C-band accelerating structure is used in section L3. The following discussions are based on the design parameters shown in Table~\ref{parameters}, which is taken from the SXFEL feasibility study report. Moreover, since the length scale in which the structural impedance is effective is much longer than that of microbunching~\cite{feasibility,Venturini1}, we may neglect the effects from the linac wakefields in the following discussions without compromising accuracy. 

\begin{table}
\caption{\label{parameters}Main beam parameters used in the microbunching instability study for the SXFEL (obtained by {\footnotesize ELEGANT}~\cite{elegant}, with laser heater turned off in simulation).}
\begin{ruledtabular}
\begin{tabular}{lllll}
 Parameter & Value \\
\hline
bunch charge (nC) & 0.5 \\ 
beam energy out of injector (MeV) & 130 \\
bunch length (FWHM) at the exit of injector (ps) & 8 \\
peak current before BC1 (A) & 60\\
beam radius (rms) before BC1 (mm) & 0.40 \\
beam energy before BC1 (MeV) & 208.4 \\
local (slice) energy spread (rms)right before BC1 (keV)& 6.5 \\
linac length up to BC1 (m)& 17.3 \\
$R_{56}$ of BC1 (mm) & 48 \\
beam energy before BC2 (MeV) & 422.0 \\
local (slice) energy spread (rms) right before BC2 (keV) & 43.4 \\
linac length up to BC2 (m)& 60.3 \\
$R_{56}$ of BC2 (mm) & 15 \\
linac length after BC2 (m)& 63.7 \\
compression ratio (BC1$\times {\rm BC2}$) & $5\times 2$ \\
beam radius (rms) before BC2 (mm) & 0.26 \\
\end{tabular}
\end{ruledtabular}
\end{table}

\subsection{Microbunching instability induced by longitudinal space-charge (LSC) effect}

\begin{figure}[htb]
   \centering
   \includegraphics*[width=60mm]{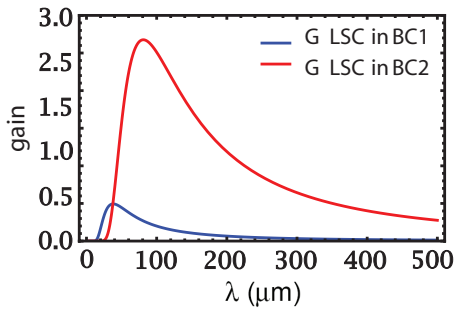}
\vskip-0.10in
   \caption{(Color) The gains of the LSC-driven microbunching instability as functions of modulation wavelength at the exits of the first (blue) compressor BC1 and second (red) bunch compressor BC2 in the SXFEL linac.}
   \label{LSCgainsep}
\end{figure}

\begin{figure}[htb]
   \centering
   \includegraphics*[width=60mm, angle=0]{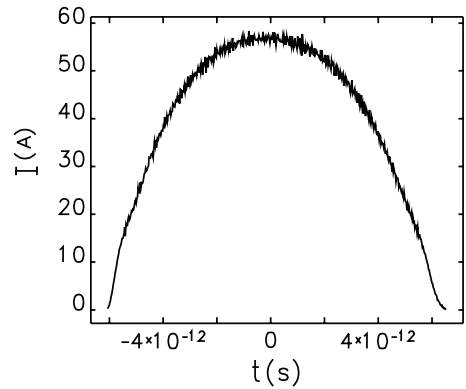}
\vskip-0.10in
   \caption{The longitudinal current profile of the input beam.}
   \label{inputcurrent}
\end{figure}

The microbunching instability for the case of linear compression has been discussed by Saldin et~al.~\cite{Saldin} phenomenologically by comparing the energy distributions before and after compression. Consider a density modulation at wavenumber $k$.  Without higher harmonics of beam current taken into account,  the gain of the instability driven by the wake fields upstream of the compressor reads~\cite{Saldin} 
\beq
G=Ck|R_{56}|\frac{I_0}{\gamma I_A}\frac{|Z_{\rm tot}(k)|}{Z_0}\exp\Bigg(-\frac{1}{2}C^2k^2R^2_{56}\frac{\sigma^2_\gamma}{\gamma^2}\Bigg).
\label{saldin}
\eeq
According to Ref.~\onlinecite{Saldin}, $\gamma$ is the nominal relativistic factor of the electron beam with rms local energy spread $\sigma_\gamma$ in front of the bunch compressor, $C=1/(1+hR_{56})$ is the compression ratio, $h$ is the linear energy chirp, $R_{56}$ is the 5-6 element of the transport matrix, $I_0$ is the initial peak current of the beam, $Z_0=377 \ \Omega$ is the free-space impedance, $Z_{\rm tot}$ is the overall impedance upstream of compressor including those of the LSC, linac wake, etc., and $I_A=17$~kA is the Alfven current. The longitudinal space-charge impedance per unit length in free space takes the form~\cite{ZHuang,Venturini}:
\begin{align}
Z_{\rm LSC}(k)&=\frac{iZ_0}{\pi kr^2_b}\Bigg[1-\frac{kr_b}{\gamma}K_1\Bigg(\frac{kr_b}{\gamma}\Bigg)\Bigg]
\nonumber \\
&\approx\left\{\begin{array}{lll}
\frac{iZ_0}{\pi kr^2_b} & \frac{kr_b}{\gamma}\gg1,\\
~&~ \\
\frac{iZ_0k}{4\pi\gamma^2}(1+2{\rm ln}\frac{\gamma}{r_bk}) & \frac{kr_b}{\gamma}\ll1,
\end{array}\right.
\label{LSCimp}
\end{align}
where $r_b$ is the radius of the beam, $K_1$ is the first-order modified Bessel function of the second kind. Based on Eqs.~(\ref{saldin}) and (\ref{LSCimp}), using the parameters output by the {\footnotesize ELEGANT} simulation starting from a beam with $\sim\pm1\%$ noise fluctuation in current and 1 -- 2 keV uncorrelated (slice) energy spread (Fig.~\ref{inputcurrent}, Fig.~\ref{LSCgainsep}),  the gains of microbunching instability in the region around the peak current induced by LSC impedance at the exits of BC1 and BC2 (first and second bunch compressors) are computed by the analytic formula and are illustrated in Fig.~\ref{LSCgainsep}. Note that in all the gain curves hereafter, the gain is the pure growth ``$G - 1$.''  The beam parameters at the exit of the SXFEL injector are prepared by {\footnotesize PARMELA}~\cite{feasibility,parmela} simulation employing one million macro-particles. The LSC impedances are computed separately in drift space and the accelerating section,

The reason why we purposely choose to use a noisy input instead of a smooth one is because the real beam is usually not ideally smooth, it always includes the ripples introduced by the shot noise fluctuation, the quantum process of field emission and the laser power jitter, etc. On the other hand, the linear theory is applied in the computation because: I. in general, the initial density modulation is very small, and II. the gain in the first compressor BC1 is small as well. Although we will point out in the next section that those assumptions are not very accurate and discrepancy exists, the linear theory does provide a good first estimate for the growth of the instability in the SXFEL linac after reviewing the results of numerical simulations (Fig.~\ref{gaincomp}). 

\begin{figure}[htb]
   \centering
   \includegraphics*[width=60mm]{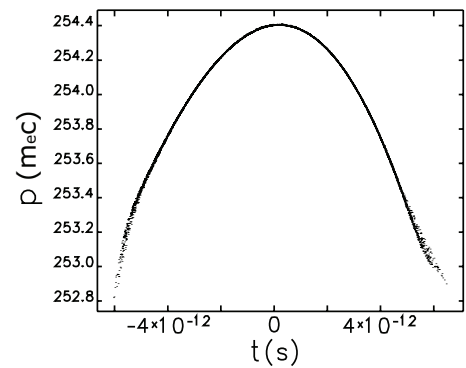}
\vskip-0.10in
   \caption{The longitudinal phase space distribution of the input beam, note that the unit of momentum is electron mass divided by the speed of light.}
   \label{inputlong}
\end{figure}

The gain curve at the exit of BC2 appears to peak at higher wavelengths, because of the larger slice energy spread after the first compression. For this reason, we will see that the longer-wavelength components in beam current lead to stronger amplification and they dominate in many cases. 
  
\subsection{Microbunching instability induced by coherent synchrotron radiation (CSR) effect in bunch compressor}

When a charged particle beam passes through a bunch compressor, coherent synchrotron radiation (CSR) can be emitted at wavelengths much shorter than the bunch itself if the density of bunch particles is modulated at those wavelengths. As we have already discussed, since the CSR effect in a bunch compressor introduces positive-feedback to the microbunching instability, 
the gain of the instability therefore rises rapidly.  

The gain of CSR-driven microbunching instability in a bunch compressor has been derived analytically~\cite{ZHuang1} in terms of beam energy, current, emittance, energy spread and chirp, as well as initial lattice and chicane parameters. Assuming a beam uniform in the $z$-direction and Gaussian in transverse and in energy distributions, the CSR-driven microbunching instability growth follows the expression~\cite{ZHuang1}:  
\begin{widetext}
\begin{eqnarray}
G_f&\approx &  \bigg |\exp\bigg [-\frac{{\bar{\sigma}}^2_\delta}{2(1+hR_{56})^2}\bigg ]+A{\bar{I}}_f\bigg [\bigg (F_0({\bar{\sigma}}_x)+\frac{1-e^{-{\bar{\sigma}}^2_x}}{2{\bar{\sigma}}^2_x}\bigg )\exp{\bigg (-\frac{{\bar{\sigma}^2_\delta}}{2(1+hR_{56})^2}\bigg )} \nonumber \\
& &+F_1(hR_{56},{\bar{\sigma}}_x,\alpha_0,\phi,{\bar{\sigma}}_\delta)\bigg ]+A^2{\bar{I}}^2_fF_0({\bar{\sigma}_x})F_2(hR_{56},{\bar{\sigma}}_x,\alpha_0,\phi,{\bar{\sigma}}_\delta)\bigg |.
\label{CSRgain}
\end{eqnarray}
\end{widetext}

As described in Ref.~\onlinecite{ZHuang1}, the first term on right side of Eq.~(\ref{CSRgain}) represents the loss of microbunching in the limit of vanishing current, the second term (linear in current) provides a one-stage amplification at low current (low gain), and the last term (quadratic in current) corresponds to the two-stage amplification at high current (high gain). The functions of $F_0$, $F_1$, and $F_2$ are defined in Ref.~\onlinecite{ZHuang}, and $\bar{\sigma}_\delta$ and $\bar{\sigma}_x$ are related to, respectively, the local (slice) energy spread and rms transverse size of the beam. 

\begin{figure}[htb]
   \centering
   \includegraphics*[width=60mm]{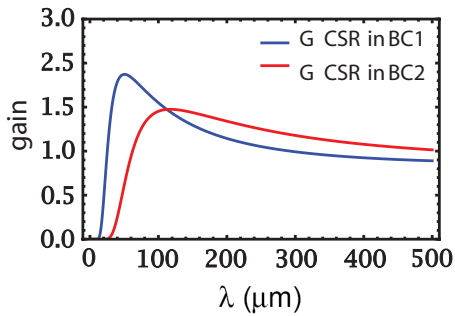}
\vskip-0.10in
   \caption{(Color) The gains of the CSR-driven microbunching instability as  functions of modulation wavelength at the exits of the first (blue) compressor BC1 and second (red) compressor BC2 in the SXFEL linac.}
   \label{CSRgainsep}
\end{figure}

The analytical computation of the gain of the CSR-driven microbunching instability in the region around the peak current is shown in Fig.~\ref{CSRgainsep}. In the calculation for the SXFEL linac, since the high-gain term in Eq.~(\ref{CSRgain}) is much smaller than the low-gain term (possibly due to the rapid increase of slice energy spread in the bunch compressor), we just ignore the contribution of the two-stage amplification. In the following section, we will show that this assumption is valid when comparison is made with simulation results.

\subsection{Discussions on final gain}

\begin{figure}[htb]
   \centering
   \includegraphics*[width=60mm]{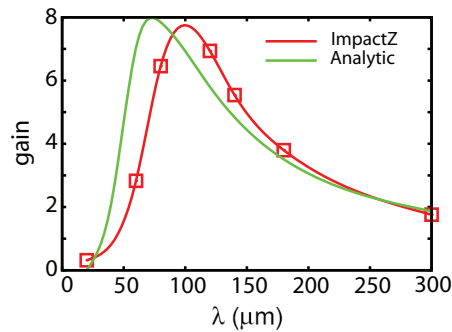}
\vskip-0.10in
   \caption{(Color) The gains of microbunching instability at the exit of the SXFEL linac section as  functions of modulation wavelength computed by analytic formulae (green) and {\footnotesize IMPACT-Z} simulation (red) based on current machine and beam parameters.}
   \label{gaincomp}
\end{figure}

{\footnotesize IMPACT-Z}~\cite{JQiang} is a three-dimensional particle-in-cell (PIC) code designed for particle tracking in an accelerator. Because it computes the electromagnetic fields based on fundamental equations, it may reproduce the dynamics of beam particles with sufficient accuracy. Another reason of choosing this code is because a PIC code performs the smoothing of the particle distribution inherently. Moreover, it is relatively fast when a large number of particles are tracked. The comparison of the gain curves at the exit of SXFEL linac obtained from analytic formulae and from {\footnotesize IMPACT-Z} simulation is shown in Fig.~\ref{gaincomp}. In the {\footnotesize IMPACT-Z} simulation, for the 8-pico-second beam out of the injector, 512 slices are used longitudinally to compute the longitudinal space charge (LSC). To obtain the data points by {\footnotesize IMPACT-Z}, initial modulation of certain wavelength is added on top of the original beam distribution, then we track the beam through the linac, and the gain is computed by comparing the Fast-Fourier transform (FFT) patterns of the beam current distribution in the vicinity of the maximal current at the entrance and the exit of the linac.    

\begin{figure}[htb]
   \centering
   \includegraphics*[width=60mm]{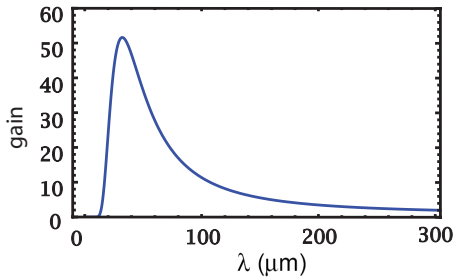}
\vskip-0.10in
   \caption{(Color) The gain curve calculated with the smoothed beam current.}
   \label{smoothgain}
\end{figure}

\begin{figure}[htb]
   \centering
   \includegraphics*[width=60mm]{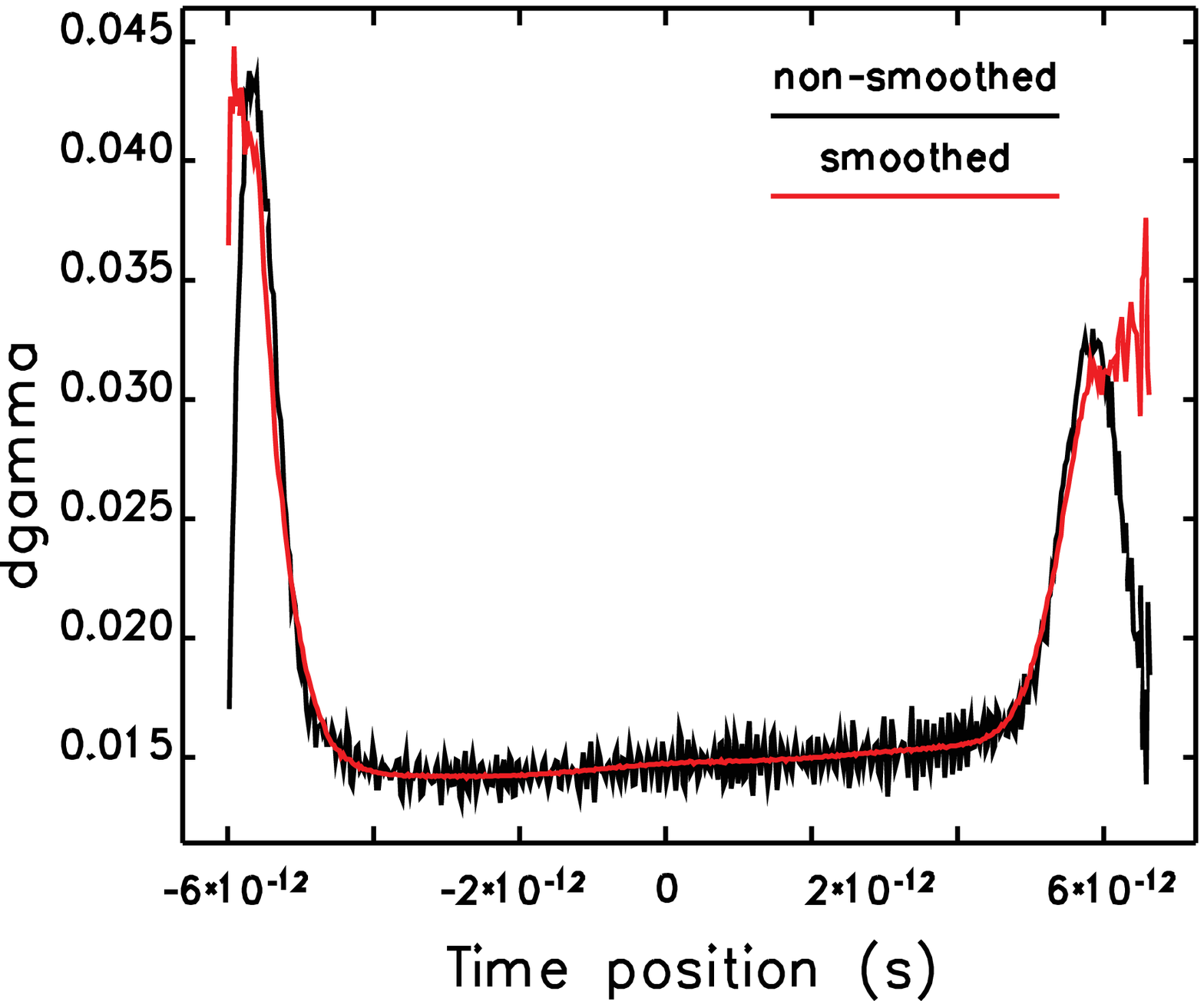}
\vskip-0.10in
   \caption{(Color) The slice energy spreads of the smoothed (red) and the non-smoothed beam (black) before BC1.}
   \label{L1dgamma}
\end{figure}

\begin{figure}[htb]
   \centering
   \includegraphics*[width=60mm]{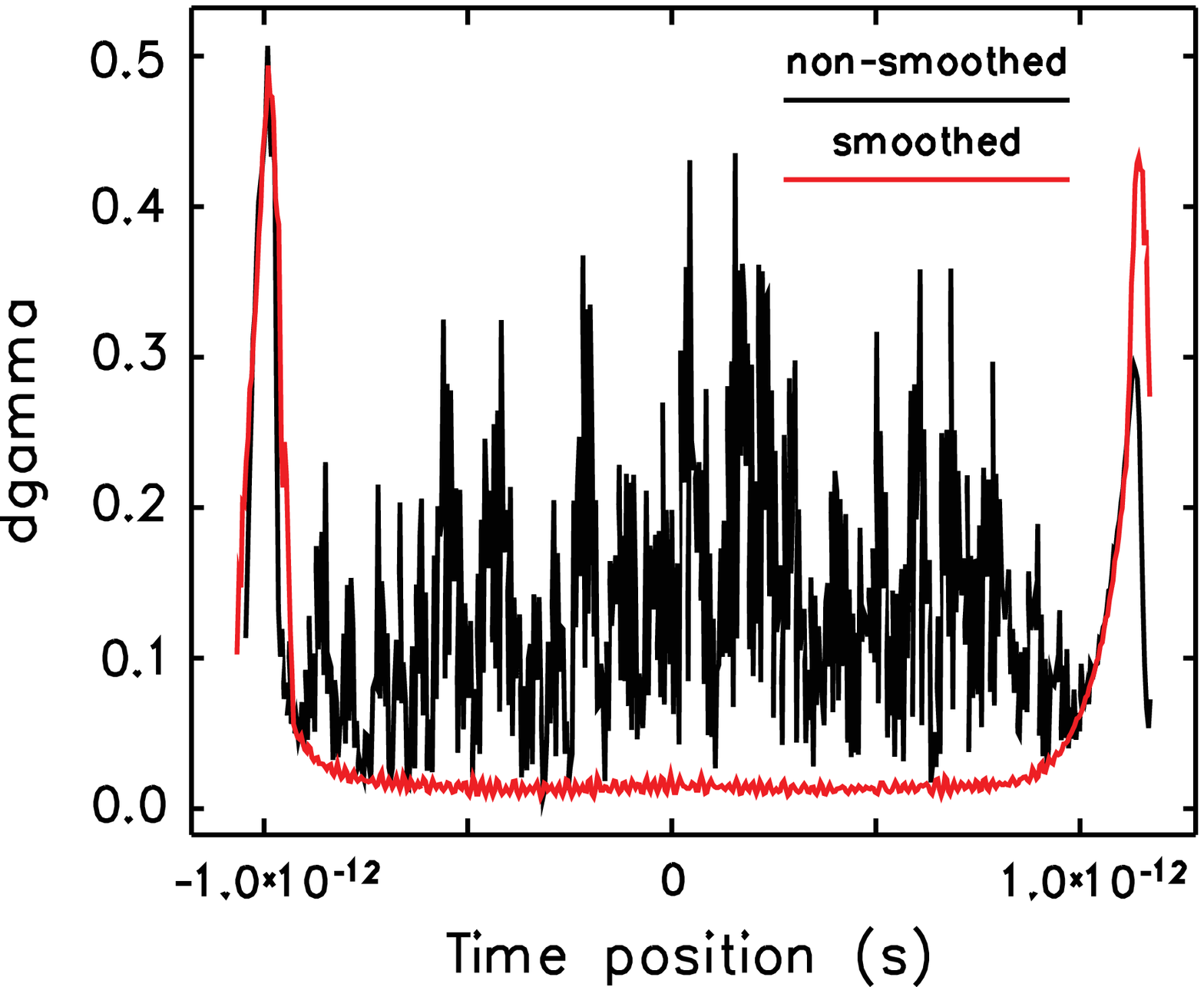}
\vskip-0.10in
   \caption{(Color) The slice energy spreads of the smoothed (red) and the non-smoothed beam (black) before BC2.}
   \label{L2dgamma}
\end{figure}

\begin{figure}[htb]
   \centering
   \includegraphics*[width=60mm]{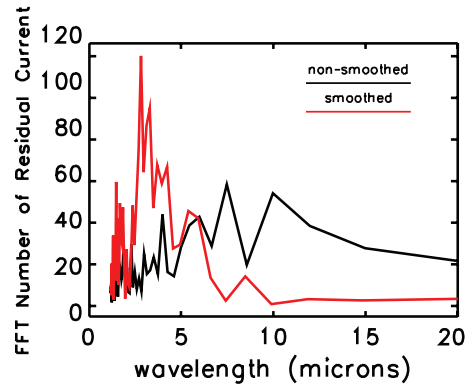}
\vskip-0.10in
   \caption{(Color) The current spectra of the smoothed (red) and non-smoothed beam (black) at the exit of the SXFEL linac, the wavelength is reduced by the factor of 10 from the initial value because of compression.}
   \label{finalcurrentFFT}
\end{figure}

As we mentioned in the previous section, there is always noise introduced inevitably when the electrons are emitted from the cathode. Therefore, in order to have a clearer picture about the effect of the initial noise on the final gain, the gain with an initially smoothed electron beam is computed as well. Our study shows that the uncorrelated (slice) energy spread before the second bunch compressor (BC2) becomes larger for an initially noisy (non-smoothed) beam input, as a result, the final gain of it is smaller than the one computed with the initially smoothed beam. Fig.~\ref{smoothgain} shows the analytically-computed gain curve of the smoothed beam; Fig.~\ref{L1dgamma} and Fig.~\ref{L2dgamma} show the uncorrelated (slice) energy spread of the smoothed and the non-smoothed beam before BC1 and BC2 by {\footnotesize ELEGANT}, respectively. All the parameters in the analytical calculation are obtained from {\footnotesize ELEGANT}, as what we do to the green curve in Fig.~\ref{gaincomp}. The term ``smoothed beam'' means the whole 6 dimensional beam distribution is smoothed by removing the noise fluctuation, which is done by the sdds command smoothDist6s~\cite{sddstoolkit}. With this command, the noise fluctuation of the beam current profile in both the longitudinal and the transverse directions, and that of the uncorrelated (slice) energy spread are all suppressed to almost zero. We can see in Fig.~\ref{smoothgain} that the peak of the curve is much higher than that obtained from the noisy beam due to the smaller slice energy spread, and the wavelength where the peak resides is also much shorter, i.e., 10 -- 20$\mu$m. However, although the total gain is small for the non-smoothed beam, the microbunching instability still can be a problem because of the significant initial noise. As the demonstration, Fig.~\ref{finalcurrentFFT} shows that the final current fluctuation of the non-smoothed beam is comparable to that of the smoothed one, which is also consistent with the analytical results (Fig.~\ref{gaincomp} and Fig.~\ref{smoothgain}) in terms of the wavelength where the peak resides. In summary, careful studies are needed to decide how much noise should be included in the computation.

\begin{figure}[htb]
   \centering
   \includegraphics*[width=60mm]{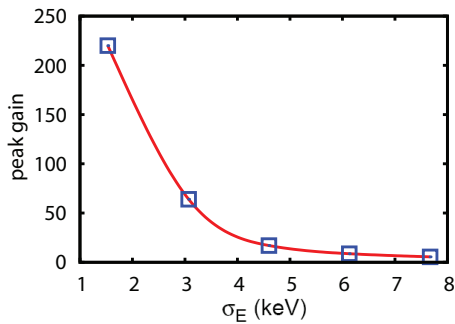}
\vskip-0.10in
   \caption{The peak gain of microbunching instability as a function of the local (slice) energy spread before the first compressor (BC1) in the SXFEL linac.}
   \label{gainslicespread}
\end{figure}

The wavelength and the magnitude of the peak appear to be slightly different in Fig.~\ref{gaincomp} for the two different methods. Possible reasons are: 

I. Beam parameters. As described, the beam parameters used in the analytic formulae are obtained by {\footnotesize ELEGANT}. Because {\footnotesize ELEGANT} and {\footnotesize IMPACT-Z} employ different algorithms to compute rf field, space charge, etc., the beam parameters calculated by the two codes could be slightly different. On the other hand, since the gain curve is very sensitive to the beam parameters such as the local (slice) energy spread, the transverse size of the beam, etc., therefore small variations of them can change the gain of the instability appreciably. As a demonstration,
we see in Fig.~\ref{gainslicespread} that, as the local energy spread rises, the peak gain of the instability falls rapidly. 
We may therefore conclude that in reality the gain can deviate from what we expect because of the possible uncertainty of the beam parameters.          

II. Numerical smoothing. Usually when a particle-in-cell (PIC) code is executed, smoothing is performed automatically for the particle density distribution in the computation of the space-charge fields because of the grid-based algorithm.  The modulation of particle density could be suppressed excessively depending upon the choice of the mesh size. For example, if the mesh size is too large, the final gain computed by {\footnotesize IMPACT-Z} can be smaller than the realistic and the modulation wavelength will be longer. Moreover, for a
formula-based simulation code like {\footnotesize ELEGANT}, smoothing is carried out by low-pass filters~\cite{Savitzky} when wakefields are computed. Therefore the same problem also exists in {\footnotesize ELEGANT}. Example relevant to noise smoothing will be shown in the next section.

III. Plasma effect. If density fluctuation exists, charged particles will move between the higher-density region and the lower-density region, which is known as space-charge (density) oscillation. It usually happens in low-energy high-intensity beams. The mechanism is similar to plasma oscillation~\cite{Jackson,Kim}. Some works have already been performed in this direction~\cite{Rosenzweig}. Preliminary investigation by solving the Vlasov and Poisson equations in the six-dimensional phase space shows that the LSC impedance (Eq.~(\ref{LSCimp})) will be modified when the beam velocity information and the space-charge oscillation frequency are included~\cite{Huangd}. In our case, the longitudinal plasma oscillation wavelength calculated based on $k_p=\sqrt{2I_e/(\gamma_0^3I_A\sigma^2_x)}$~\cite{HuangKim} is about 120 m, where $k_p$ is the wave number of the oscillation, $I_e$ is the electron peak current, $\gamma_0$ is the electron energy, $\sigma_x$ is the horizontal beam size and $I_A\approx17$ kA is the Alfv\'en current. Because the plasma wavelength is much larger than the length of the first acceleration section (L1) ($\approx$ 6 m) before BC1, we thus do not need to pay too much attention on the plasma effect at this point. However, the plasma effect in the injector must be investigated because of the low electron energy.      

\section{critical issues in the study}

In our study, some new issues are found to have profound effects on the final results of the microbunching instability and they need to be included in the computation. These issues include higher harmonics of beam current, transverse-longitudinal coupling in a bunch compressor, and the method of noise smoothing. In this section, we will discuss these issues one by one in detail.

\subsection{Higher order harmonics of beam current} 

\begin{figure}[htb]
   \centering
   \includegraphics*[width=55mm,angle=0]{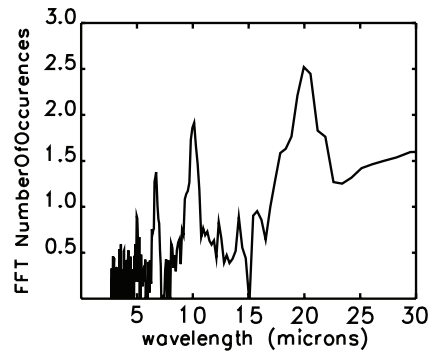}
\vskip-0.10in
   \caption{The Fast Fourier Transform (FFT) spectrum of the beam current with initial modulation wavelength of 100 $\mu$m at the exit of BC1, where 20 $\mu$m in this figure corresponds to the initial 100 $\mu $m because of compression ratio of 5.}
   \label{harmonics}
\end{figure}

\begin{figure}[htb]
   \centering
   \includegraphics*[width=55mm,angle=0]{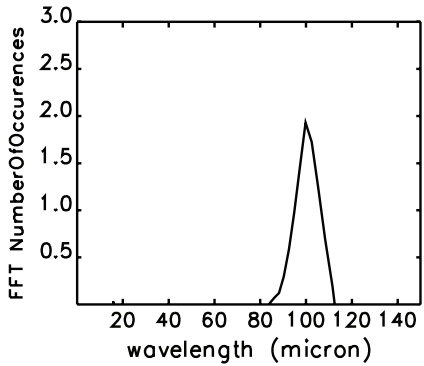}
\vskip-0.10in
   \caption{The Fast Fourier Transform (FFT) spectrum of the beam current with initial modulation wavelength of 100 $\mu$m at the exit of the injector.}
   \label{initcurFFT}
\end{figure}

Equation~(\ref{saldin}) only takes into account the gain of the fundamental mode of beam current. In other words, the gain curve obtained from it merely describes the behavior of the fundamental mode and nothing else, e.g., Fig.~\ref{gaincomp} and Fig.~\ref{smoothgain} in this paper. The linear theory does not include the higher order harmonics, which may exist all the time simultaneously. However, in our simulations, higher-order harmonics of beam current are also observed. We therefore believe that, in certain cases, higher harmonics should be included in the computation of the gain. Preliminary studies tell us that the gain of the higher harmonics depends not only on the peak current of the beam, but also on the modulation depth. Thus the inclusion of the higher-harmonics becomes a bit more complicated than the fundamental mode. As an example, Fig.~\ref{harmonics} shows the excitation of higher harmonics simulated by {\footnotesize ELEGANT} after the first bunch compressor (BC1) with initial modulation wavelength of 100 $\mu$m (corresponding to 20 $\mu$m in Fig.~\ref{harmonics}) which is in the vicinity of the peak (Fig.~\ref{gaincomp}). The initial modulation depth is 5\% and The Fast Fourier Transform (FFT) spectrum of the initial beam current is illustrated in Fig.~\ref{initcurFFT}. We see in Fig.~\ref{harmonics} that, besides the peak gain of the fundamental residing around 20 $\mu$m, there are also the first, second, and third harmonics excited at, respectively, 10, 6.7, and 5 $\mu$m. Similar excitations are observed as well at other initial modulation wavelengths such as $80~\mu$m, $60~\mu$m, and $35~\mu$um. Thus we need to keep in mind that higher-order of the initial modulation can also be strongly excited and the linear growth theory should be extended.

\subsection{Transverse-longitudinal coupling}

One important topic in microbunching-instability study is noise. Like the self-amplified spontaneous emission (SASE) process, micro-bunches can be generated from shot noise. If the beam is ideally smoothed in all directions, i.e., without any noise, fluctuation, modulation, etc., microbunching instability can hardly be excited. To reduce microbunching, there are ways to smooth the longitudinal current profile from the cathode by improving the temporal stability of driving laser. However, our investigation indicates that the density noise in the transverse direction can also introduce roughness to the longitudinal beam density distribution as a result of transverse-longitudinal coupling, for example, the $R_{51}$ elements in the transfer matrices of the dipoles in a bunch compressor.
\begin{figure}[htb]
   \centering
   \includegraphics*[width=60mm,angle=0]{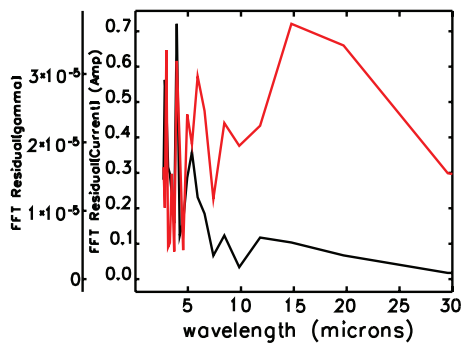}
\vskip-0.10in
   \caption{(Color) The current (black) and energy (red) spectra of a beam smoothed in all directions at the exit of BC1 with the laser heater turned off.}
   \label{smoothall}
\end{figure} 

\begin{figure}[htb]
   \centering
   \includegraphics*[width=60mm,angle=0]{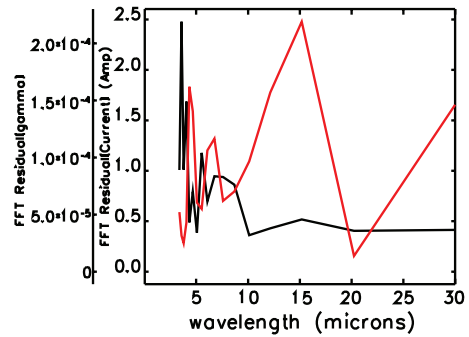}
\vskip-0.10in
   \caption{(Color) The current (black) and energy (red) spectra of a beam smoothed only in longitudinal direction at the exit of BC1 with the laser heater turned off.}
   \label{smoothlong}
\end{figure}

Assuming a beam smoothed in all directions, and another beam with $\thicksim5\%$ noise level in transverse but smoothed longitudinally, Figs.~\ref{smoothall} and~\ref{smoothlong} show the current and energy spectra of the beams at the exit of the first chicane (BC1) without laser heating. In the figures, one can see that the levels of density and energy fluctuation (modulation) of the 3-D smoothed beam are undoubtedly smaller than those when the beam is smoothed only longitudinally.   

\begin{figure}[htb]
   \centering
   \includegraphics*[width=80mm,angle=0]{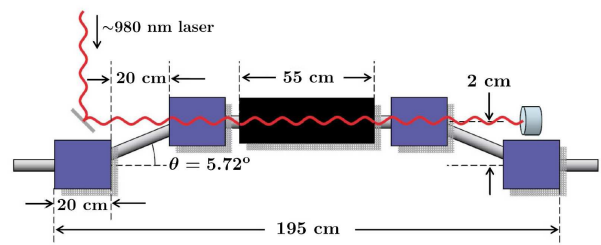}
\vskip-0.10in
   \caption{The layout of the SXFEL laser heater at 130 MeV.}
   \label{laserheater}
\end{figure}

\begin{figure}[htb]
   \centering
   \includegraphics*[width=60mm]{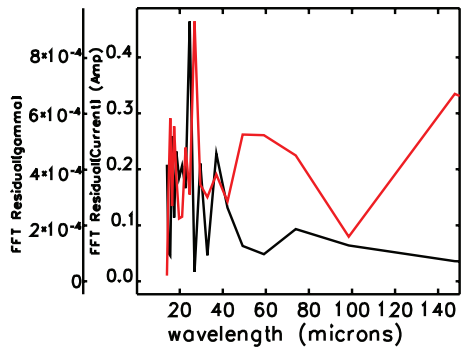}
\vskip-0.10in
   \caption{(Color) The current (black) and energy (red) spectra of a beam longitudinally smoothed but transversely noisy at the first BPM right after the laser heater in the SXFEL linac, with the laser heater turned on in the simulation.}
   \label{LH}
\end{figure}

\begin{figure}[htb]
   \centering
   \includegraphics*[width=60mm]{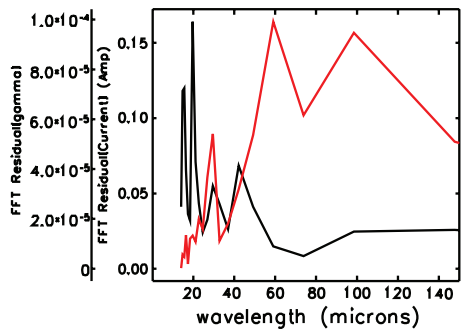}
\vskip-0.10in
   \caption{(Color) The current (black) and energy (red) spectra of a beam longitudinally smoothed but transversely noisy at the first BPM right after the location of the laser heater in the SXFEL linac, with the laser heater substituted by a drift in the simulation.}
   \label{noLH}
\end{figure}

Nowadays, a laser heater has been a common device to suppress the microbunching instability~\cite{ZHuang} in the linac of a FEL facility. As an example,  Fig.~\ref{laserheater} shows the structure of the SXFEL laser heater including an injection laser, an undulator, and a chicane-type bunch compressor. Because of the transverse-longitudinal coupling, some extra noises are introduced into the longitudinal direction and the amplitudes of the modulations are also amplified when beam passing through the laser heater. Figures~\ref{LH} and~\ref{noLH} show the current and energy spectra of a beam longitudinally smoothed but transversely noisy at the location right after the designed location of the laser heater with and without the laser heater included in the simulations. We can clearly see that the laser heater introduces extra peaks at various wavelengths and also increases the amplitudes of the current and energy modulation, which will compromise the beam quality thereafter. For this reason, the $R_{51}$ element of the magnetic chicane should be revisited with more care. The parameters of the SXFEL laser heater used in the simulation are: length 0.55 m, periods 10, undulator magnetic peak field 0.31 T, laser peak power 0.80 MW, laser spot size at waist 0.30 mm.

\subsection{Noise smoothing}

When computing wakefields, noise smoothing is always performed by the simulation code to obtain the beam distribution. For example, in {\footnotesize ELEGANT}, the Savitzky-Golay filter is used to smooth the longitudinal beam current profile (thus reducing the noise) in the course of computing various kinds of wakefields. Because the method of smoothing has a strong influence on the final results, it must be investigated with care.

The Savitzky-Golay (S-G) smoothing filter is a filter that performs a local polynomial regression (of degree $k$) on a series of values (of at least $k+1$ points which are treated as being equally spaced in the series) to determine the smoothed value for each point~\cite{wiki}. This filtering algorithm was first described in 1964 by Abraham Savitzky and Marcel J. E. Golay~\cite{Savitzky}. They proposed a method of data smoothing based on local least-square polynomial approximation and showed that fitting a polynomial to a set of input samples and then evaluating the resulting polynomial at a single point within the approximation interval is equivalent to the discrete convolution with a fixed impulse response. The low-pass filters obtained by this method are widely known as Savitzky-Golay filters. In general, there are two key parameters in this method, one is the order, the other is the half-width. The former represents the highest order in the polynomial fit, while the latter corresponds to the number data points needed to be sampled in the smoothing. 

\begin{figure}[htb]
   \centering
   \includegraphics*[width=55mm,angle=0]{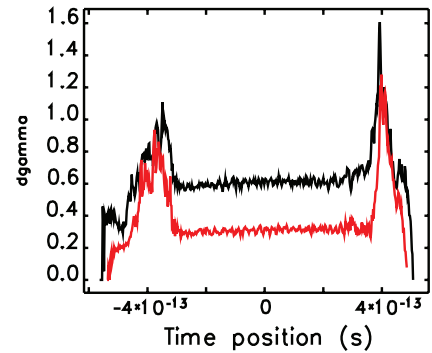}
\vskip-0.10in
   \caption{(Color) The local (slice) energy spread at the exit of the SXFEL linac with the laser heater turned on. The S-G filter half-width is set at 1 (black) and 10 (red).}
   \label{SGdgamma}
\end{figure} 
In our study, we vary the half width of the S-G filter in the {\footnotesize ELEGANT} input, and discover that the difference between the output slice energy spreads with different S-G half-widths is significant (Fig.~\ref{SGdgamma}). This is understandable because stronger suppression of noise results in stronger reduction of local energy spread (excessive smoothing). Therefore here comes the problem: how to differentiate the real noise from the numerical noise in simulation? We suggest to use the S-G filter with caution (e.g., using filter half-width~$\leqslant5$) and to use as many macro-particles as possible within the capability of hardware. In our {\footnotesize ELEGANT} simulations here, we fix the S-G filter half-width at 1 and employ 1 million macro-particles. 

\section{conclusions}
The study of microbunching instability in the SXFEL linac is performed in detail as a classic example. The potential problems and some new effects are uncovered including the initial beam distribution, the higher order modes in the noise amplification, the transverse-longitudinal coupling and the plasma effect in the computation of LSC. More investigations are needed to obtain deeper insights of those problems. 

Computations using both analytic expressions and numerical simulations show that the gain of the microbunching instability indicates large discrepancy between the noisy and the smoothed beam input. Since in the real case, there is always noise introduced by the shot noise, the laser power jitters, etc., one should consider how much noise to be included in the initial input. On the other hand, our work shows that in terms of the final density/energy fluctuation (or the bunching factor), the difference between the noisy and the smoothed input is not large.

A particle-in-cell (PIC) code appears to be a better tool for the study of microbunching instability because of its intrinsic advantage\,---\,the electromagnetic fields are computed by solving fundamental physical equations, i.e., the Poisson solver. As a result, a PIC code can be better in revealing the detailed physical process of the instability than a code based on analytic formulae. However, like all the other codes, the noise-smoothing method must be investigated to see whether it will introduce extra uncertainty. The challenging problem is how to separate and discard the numerical noise from the real one. Currently, we tend to preserve the information in the beam as much as possible by employing the least smoothing.

The higher-order harmonics of beam current that revises the linear theory in the estimate of the gain should also be included in the development of the analytic formulae, especially in the vicinity of the peak gain. The transverse-longitudinal coupling in a magnetic chicane will transport shot noise from the transverse dimension to the longitudinal and needs to be considered in the design of the laser heater. In this sense, we believe that the microbunching instability in an electron linac cannot be completely avoided unless the beam is ideally smooth in the whole 6 dimensional phase space.

Because both the analytic and numerical methods exhibit limitations in the estimate of the microbunching instability, systematic experimental measurements are desired to provide solid and full understanding of the physical essence. Work is ongoing at the Shanghai Institute of Applied Physics (SINAP) to prepare these experiments.

\begin{acknowledgments}
We wish to acknowledge the help of Dr.~Yuantao Ding from SLAC National Accelerator Laboratory for the suggestions of {\footnotesize ELEGANT} simulation, and the discussions with Dr.~Ji Qiang from Lawrence Berkeley National Laboratory on the {\footnotesize IMPACT-T/Z} particle tracking code, and many colleagues in SINAP for discussions on the analytical and simulation results. The work is supported by National Science Foundation of China (NSFC), grant No. 11275253. 

\end{acknowledgments}


\begin{thebibliography}{9}   
\bibitem{Saldin}
E.L. Saldin, E.A. Schneidmiller, M.V. Yurkov, NIMA \textbf{483} (2002) 516-520

\bibitem{Heifets}
S. Heifets, G. Stupakov, S. Krinsky, Phys. Rev. ST Accel. Beams \textbf{5}, 064401 (2002)

\bibitem{ZHuang1}
Zhirong Huang and Kwang-Je Kim, Phys. Rev. ST Accel. Beams \textbf{5}, 074401 (2002) 

\bibitem{ZHuang}
Z. Huang, M. Borland, P. Emma, et~al., Phys, Rev. ST Accel. Beams \textbf{7}, 074401 (2004)


\bibitem{Savitzky}
A. Savitzky, M.J.E. Golay, Anal. Chem. \textbf{36} (8): 1627-1639

\bibitem{feasibility}
SXFEL feasibility study report, Nov. 28th, 2011

\bibitem{Venturini1}
M. Venturini, R. Warnock, and A. Zholents, Phys. Rev. ST Accel. Beams \textbf{10}, 054403 (2007).

\bibitem{elegant}
M. Borland, Advance Photon Source LS-287, Sep. 2000

\bibitem{Venturini}
Marco Venturini, Phys. Rev. ST Accel. Beams \textbf{11}, 034401 (2008)

\bibitem{parmela}
Lloyd Young and James Billen, Proc. of PAC03, May 2003, Portland, United States 


\bibitem{JQiang}
J. Qiang, Robert D. Ryne, Salman Habib, et~al., J. of Comp. Phys. \textbf{163}, 434-451 (2000)

\bibitem{sddstoolkit}
M. Borland, L. Emery, H.Shang and R. Soliday. User's Guide for SDDS Tookit Version 1.30, June 23, 2009

\bibitem{Jackson}
J.D. Jackson, J. Nucl. Energy, Part C: Plasma Physics, 1960, vol \textbf{1}, 171-189

\bibitem{Kim}
K.-J. Kim and R.R. Lindberg, Proc. of FEL11, Aug. 2011, Shanghai, China


\bibitem{Rosenzweig}
J. Rosenzweig, C. Pellegrini, L. Serafini, et~al., DESY Report No. TESLA-FEL-96-15, 1996.

\bibitem{Huangd}
Dazhang Huang, King Yuen Ng, Qiang Gu, Proc. of LINAC12, Sep. 2012, Tel-Aviv, Israel

\bibitem{HuangKim}
Zhirong Huang and Kwang-Je Kim, Phys. Rev. ST Accel. Beams \textbf{10}, 034801 (2007). 

\bibitem{wiki}
http://en.wikipedia.org/wiki/Savitzky-Golay\_smoothing\_filter


\end{thebibliography}
\end{document}